\documentclass[prb,reprint,notitlepage]{revtex4-1}

\usepackage{amsmath, amssymb, amsthm}  
\usepackage{latexsym, graphicx}

\usepackage{float}           % use in \begin{figure}[H] to enforce position
\theoremstyle{plain}
\newtheorem*{xtheorem}{Theorem}      	      
\begin{document}

\title{Geometric Diagram\\ for Relativistic Addition of Velocities}%[Relativistic diagram]

\author{Jerzy Kocik}
 
\affiliation{Department of Mathematics, Southern Illinois University, Carbondale, IL62901}

\email{jkocik{@}siu.edu}

%\date{}

\begin{abstract}
A geometric diagram that allows one to visualize the Poincar\'e formula for relativistic addition 
of velocities in one dimension is presented.  
An analogous diagram representing 
the angle sum formula for trigonometric tangent is given.
\footnote{{\it Am. J. Phys.}, Aug 2012, {\bf 80}  (8),  p. 737.}
\end{abstract}

\maketitle

\section{Relativistic diagram}

If James runs atop a train and his velocity with respect to the train is $a$, 
while the train has velocity $b$ with respect to ground, 
what is the resulting velocity of James with respect to  ground?  
Before 1905, we would answer as Galileo would: $a+b$.  
But now we know better: his velocity is 
\begin{equation}
a \oplus b  =   \frac{a+b}{1+ab}\,.
\end{equation}
This elegant expression 
was discovered by Henri Poincar\'e \cite{Po} and constitutes 
the relativistic velocity-addition law in one dimension 
(for parallel velocities). 
Velocities are expressed here in natural units, 
in which the speed of light (unattainable for objects with mass) is 1.

%================= FIGURE 1
\begin{figure}[h]
\centering
\includegraphics[scale=.6]{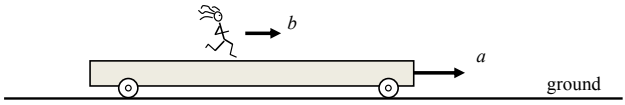}
\caption{\small Velocity a of the big platform a is measured with respect to the ground.  Velocity b of the runner -- with respect to the platform.  What is the velocity $a\oplus b$ of the runner with respect to the ground?}
\label{fig:train}
\end{figure}

It is fun to see how such a simple algebraic expression takes care 
of different physical situations:
$$
\begin{array}{rcl}
\hbox{\scriptsize\sf  James cannot exceed the speed of light:}
&\quad\!\!\!\!&        		 
\frac{1}{2}\oplus \frac{1}{2} =  \frac{\frac{1}{2} + \frac{1}{2}} {1+\frac{1}{2}\frac{1}{2}} = \frac{4}{5} <1
\\[10pt]
\hbox{\scriptsize\sf Even if he tries really hard:}
&&
\frac{4}{5}\oplus \frac{5}{6} =  \frac{\frac{4}{5} + \frac{5}{6}} {1+\frac{4}{5}\frac{5}{6}} = \frac{49}{50} <1
\\[10pt]
\hbox{\scriptsize\sf or flashing light while running}
&&
v\oplus1 =  \frac{v+1}{1+v\cdot 1} = 1
\\[10pt]
\hbox{\scriptsize\sf In both directions:}
&&
v\oplus -1 =  \frac{v-1}{1+v\cdot(-1)} = -1
\\[10pt]
\hbox{\scriptsize\sf The desperate extreme addition:}                 	
&&
1\oplus 1 =  \frac{1+1}{1+1\cdot 1} = 1
\end{array}
$$

The purpose of this note is to present a simple geometric diagram (Figure 2) 
that allows one to visualize the Poincar\'e formula, 
and to perform the addition in a purely geometrical way.

%================= FIGURE 2
\begin{figure}[h]
\centering
\includegraphics[scale=.85]{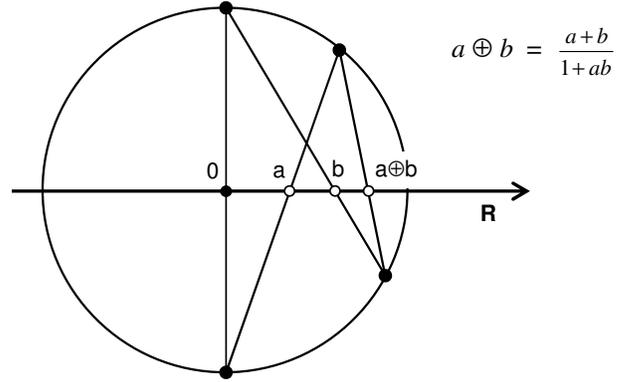}
\caption{\small Visualization of the formula for  relativistic addition of velocities}
\label{fig:symbols}
\end{figure}

In the figure, velocities are represented as points $a$ and $b$ on the real line $\mathbb R$. 
The line is embedded in $\mathbb R^2$ and the unit circle at the origin is added. 
Each velocity determines a point on the circle: namely the intersection 
points with the line from $(0,1)$ through $a$, and with the line from $(0,-1)$ through $b$ (or vice versa). 

\begin{xtheorem}
The point on ${\mathbb R}$ that belongs to the line joining the two points just constructed 
represents the relativistic composition $a\oplus b$.   
\end{xtheorem}

%------------------  begin proof
\noindent 
{\bf Proof:}  Let A and B be the points on the circle determined by 
velocities $a$ and $b$ (for economy, $a$ and $b$ denote both points 
on $\mathbb R$ and the corresponding values). 
First, we find their coordinates $(x,y)\in \mathbb R^2$.  
The result, presented in the Fig.~3, may be obtained as follows.

%================= FIGURE 3
\begin{figure}[h]
\centering
\includegraphics[scale=.8]{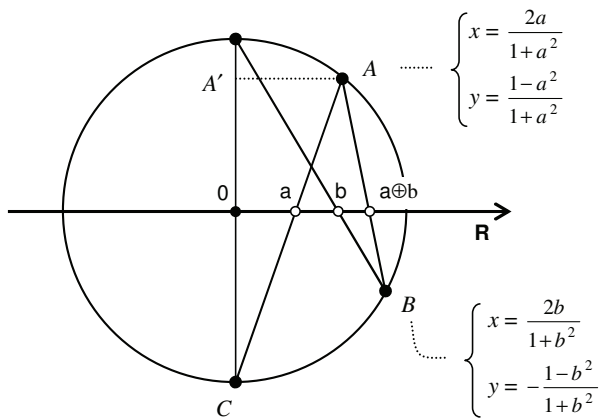}
\caption{\small  Coordinates of points A and B on the circle.}
\label{fig:symbols}
\end{figure}

Use similarity of triangles $(A, A', C)$ and $(a, 0, C)$ to
determine the coordinate value $x$ of point $A$ :
\begin{equation}
\label{eq:ratio}
\frac{a}{1} = \frac{x}{1+y}
\end{equation}
Squaring the above expression gives:
$$
a^2 =  \frac{x^2} {(1+y)^2}   =  \frac{1-y^2} {(1+y)^2}  
        =  \frac{(1-y)(1+y)} {(1+y)^2}  = \frac{1-y} {1+y}  
$$
from which one readily extracts $y$:
\begin{equation}
y =      \frac{1-a^2}{1+a^2}\,  .
\end{equation}
Applying this to (\ref{eq:ratio}) one gets
\begin{equation}
x =  \frac{2a}{1+a^2}\,  .
\end{equation}
The coordinates for point $B$ follow from symmetry.  
With these two points, $A$ and $B$, in hand, one constructs the line through them:
$$
y \ = \ \frac{1+ab}{a-b} x - \frac{a+b}{a-b}  
\quad\hbox{ or } \quad       
 (1+ab) x + (b-a)y = a+b
$$

Now, by substituting  $y = 0$, we get the point of intersection 
with the horizontal axis:
$$
(x,\,y) \ = \ \ \left( \frac{a+b}{1+ab} , \, 0 \right) \, ,  
$$
which proves the theorem.  $\square$
\\

%--------------------------------------------------------------------------
\section{Experiments}  

The diagrammatic interpretation allows one to visualize 
basic algebraic features of the relativistic addition formula.  
A few situations are shown in Fig.~4. 
Some extreme cases are illustrated in the bottom row.
An interactive applet for more explorations may be 
found on the author's web page \cite{jk1}  
(one can examine there what happens in the ``forbidden" case of $1\oplus(-1)$).

%================= FIGURE 4
\begin{figure}[h]
\centering
\includegraphics[scale=.65]{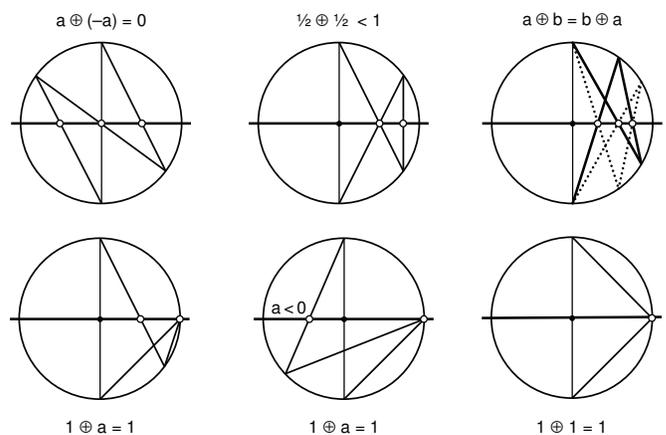}
\caption{\small Various experiments in algebra done with geometry.}
\label{fig:f4}
\end{figure}

One of the speculative theories of physics is the existence of tachyons, 
hypothetical particles that move faster than light.\cite{B,F} 
According to the theory, one of its counterintuitive properties is that, 
unlike ordinary particles, a tachyon slows down as its energy increases. 
Tachyons are forbidden from slowing down below the speed of light, 
as regular matter cannot speed up above this barrier.  
(In both cases infinite energy would be required to reach 
the speed of light from either side).  
One can see this with the aid of our diagram (see Fig. 5): 
adding a small velocity to a tachyon slows it down.  
And subtracting --- speeds it up!

%================= FIGURE 5
\begin{figure}[h]
\centering
\includegraphics[scale=.75]{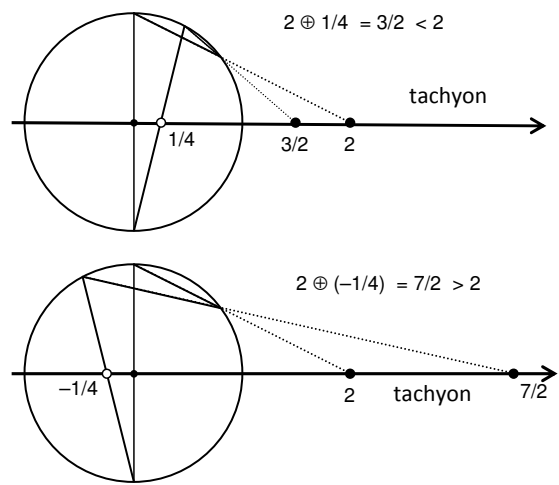}
\caption{\small Experiments with tachyons.}
\label{fig:f5s}
\end{figure}

We leave it to the reader to draw a diagram showing 
that adding two tachyonic velocities would result a subluminal
speed, normal for the common matter.

%------------------------------------------------------------------------------------------
%\newpage

\section{Trigonometric tangent}

The Poincar\'e formula is similar to the addition formula for tangents 
in regular trigonometry:  
$$                
\tan{(\alpha+\beta)} \ = \ \frac{\tan\alpha+\tan\beta}{1-\tan\alpha\tan\beta}
\qquad
a \oplus b \ = \ \frac{a + b}{1-ab}
$$
for economy, we use the same symbol $\oplus$ to denote  
the tangent addition.

Figure 6 shows how one may adjust the previous idea to 
the trigonometric case --- namely, the regular circle must be 
replaced by a ``hyperbolic circle".
Figures 7 and 8 show a few examples.   
It is intriguing if not paradoxical that one needs a circle 
to make a construction for hyperbolic geometry 
and a hyperbola for the geometry of circle...

%================= FIGURE 6
\begin{figure}[h]
\centering
\includegraphics[scale=.9]{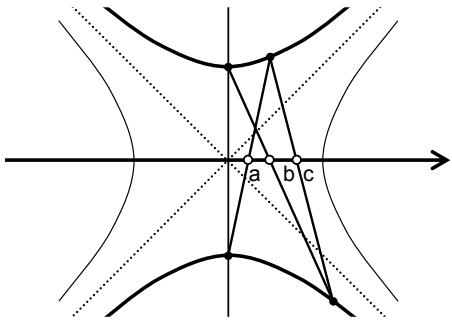}
\caption{\small Tangent-like addition}
\label{fig:symbols}
\end{figure}

The last comment is about the analogy between the trigonometric tangent and its hyperbolic version.
Actually, one may continuously deform one into the other.  Deform the circle in Figure 1 into an ellipse through 
points $(0, \pm1)$ and with a wider horizontal size while maintaining the geometrically defined addition.
 Eventually the ellipse turns into a pair of parallel lines, $y=\pm1$ for which the geometric addition 
coincides with regular arithmetic addition. Increasing further the eccentricity will 
result in the hyperbolic case. 
The transformation may be executed by changing  continuously the value of $a$ from $1$  to $-1$
in the quadratic $ax^2 + y^2=1$.

%================= FIGURE 7
\begin{figure}[h]
\centering
\includegraphics[scale=.7]{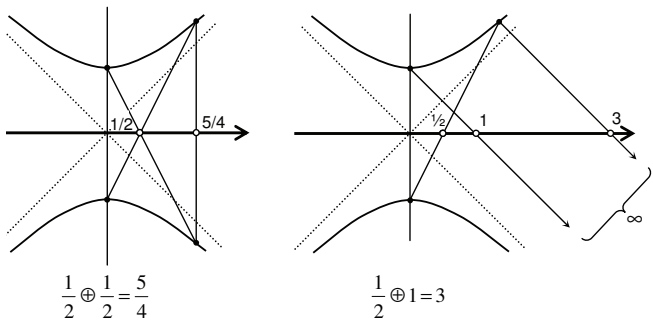}
\caption{\small  Interesting cases of the tangent addition formula}
\label{fig:symbols}
\end{figure}

%================= FIGURE 7
\begin{figure}[h]
\centering
\includegraphics[scale=.7]{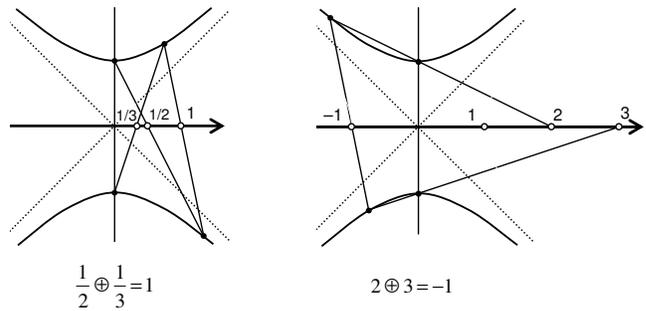}
\caption{\small  More tangent additions}
\label{fig:symbols}
\end{figure}

%-----------------------------------------------------
\section{Historical note}

The relativistic velocity addition formula  first appears in the letter of Poincar\'e to Lorentz, roughly in May 1905, 
and was presented to the French Academy of Science in 6 June of the same year.  
Basic ingredients of relativity principle were presented by Poincar\'e at the world scientific conference 
in  St Louis in September 1904.  
Poincar\'e also noted that the expression $d^2  =  x^2 + y^2 + z^2 -  c^2t^2$
defines a relativistic invariant.  Although he used 4-vectors, 
Poincar\'e did not pursue the geometric consequences. 
It was Hermann Minkowski who followed up on this idea in 1907. 
Albert Einstein decisively removed the concept of ether and simplified the derivation of 
Lorenz' transformation rules in 1905.  
For more on history of relativity see Refs~\onlinecite{Z,K,Pa, Cer}.
For a different geometrization of relativistic velocity addition see Ref.~\onlinecite{Bre}.

\section{\bf Remark} 
A lightly shorter version of this text appeared under the same name in 
{it Am. J. Phys.}, Aug 2012, {\bf 80}  (8),  p. 737.  
The part on trigonometry was removed due to a strong suggestion of one of the reviewers.

%Bibliography-------------------------------------------------------------------------------

~\\


\begin{thebibliography}{9}

\bibitem % 
{Po} Henri Poincar\'e, 
Letter to Hendrik Lorenz, ca. May 1905, available 
at http://www.univ-nancy2.fr/poincare/chp/text/lorentz4.xml.

\bibitem % 
{jk1}Jerzy Kocik, 
        Interactive diagram for relativistic velocity addition, 
http:/\!/Lagrange.math.siu.edu/Kocik/geometry\!\! /geometry.htm

\bibitem 
{B}   Olexa-Myron Bilaniuk,  Vijay K. Deshpande and  E. C. George Sudarshan,  
        ``Meta Relativity,''  Am. J. Phy.  {\bf 30},  718--723  (1962).

\bibitem 
{F} Gerald Feinberg,  ``Possibility of Faster-Than-Light Particles,'' 
      Phys.  Rev.  {\bf 159},   1089--1105  (1967).

\bibitem 
{Z} Elie Zahar,  ``Poincar\'e's Independent Discovery of the relativity principle," 
    Fundamenta Scientiae {\bf 4}, 147--175 (1983).

\bibitem 
{K} Gobind Hemraj Keswani, ``Origin and Concept of Relativity, Parts I, II, III," 
     Brit. J. Phil. Sci. {\bf 15-17}, (1965-6).  

\bibitem 
{Pa} Abraham Pais,  {\it Subtle is the Lord: The Science and the Life of Albert Einstein},
        (New York, Oxford University Press, 1982).

\bibitem 
{Cer} Roger Cerf,  ``Dismissing renewed attempts to deny Einstein the discovery of special relativity,'' 
                            Am. J. Phys. {\bf 74},  818--824 (2006).
\bibitem 
{Bre} Robert W. Brehme, ``Geometrization of the relativistic velocity addition formula,''
             Am. J. Phys.  {\bf 37}, 360--363   (1969). 


\end{thebibliography}
\end{document}